# Effect of L2$_1$ and XA ordering on phase stability, half-metallicity and magnetism of Co$_2$FeAl Heusler Alloy: GGA and GGA+U approach

Aquil Ahmad*[a], S.K. Srivastava[a], and A.K. Das[†a]

[a]Department of Physics, Indian Institute of Technology, Kharagpur, West Bengal, India-721302

*Email: aquil@phy.itkgp.ac.in

**Abstract:** The generalized gradient approximation (GGA) scheme in the first-principles calculations are used to study the effect of L2$_1$ and XA ordering on the phase stability, half-metallicity and magnetism of Co$_2$FeAl (CFA) Heusler alloy. Various possible hypothetical structures: L2$_1$-I, L2$_1$-II, XA-I, and XA-II were prepared under the conventional L2$_1$ and inverse XA phases by altering the atomic occupancies at their Wyckoff sites. It is found that the XA-II phase of CFA is the most stable phase energetically among all the structures. The electronic structure calculations without U show the presence of half-metallic (HM) ground state only in L2$_1$-1 structure and the other structures are found to be metallic. However, the electronic structures of CFA are significantly modified in the presence of U, although the total magnetic moments per cell remained the same and consistent with the Slater-Pauling (SP) rule. The metallic ground states of CFA in L2$_1$-II and XA-II structures are converted into the half-metallic ground states in presence of U but remained the same (metallic) in XA-I structure. The results indicate that the electronic structures are not only dependent on the L2$_1$ and XA ordering of the atoms but also depend on the choice of U values. So experiments may only verify the superiority of GGA+U to GGA.

## 1. Introduction

Heusler alloys (HAs) are the materials of interest of the scientific community for decades. The HAs family comprises three subfamilies: the half Heusler alloys, the full-Heusler (L2$_1$ structure) alloys, and the inverse Heusler (XA structure) alloys. The subfamilies of HAs were well described somewhere else [1]. They showed the diverse physical properties such as half-metallicity, topological insulators, spin gapless semiconductors (SGS) including the superconductivity, shape memory effect induced by the magnetic field and the magnetocaloric effect [2-16]. Co$_2$-based Heusler compounds were synthesized first time in its ideal L2$_1$ structure in 1970 [17]. They exhibit a wide range of magnetic properties with record magnetic moments up to 6.5 µ$_B$/f.u. and Curie temperature up to 1261 K [18-20]. It was theoretically predicted that the Co$_2$-based Heusler compounds should behave as a half-metal even at room temperature [21, 22]. Enhanced performance has been reported based on the theoretical design and the synthesis protocol of HAs, which indicates that their physical properties are strongly dependent on their structural orderness/disorderness [5-8, 23]. F. Dahmane et al. [24] studied the L2$_1$ (Cu$_2$MnAl prototype) and XA (Hg$_2$CuTi prototype) ordering effect on the phase stability of Fe$_2$XAl (X= Cr, Mn, Ni) Heusler compounds. They found that the L2$_1$ phase of Fe$_2$CrAl and Fe$_2$MnAl are more stable than the XA phase at the equilibrium volume. Xiaotian Wang et al. [25], reported the effect of L2$_1$ and XA ordering on Hafnium-based full-Heusler alloys and found that all of them were likely to stabilize in L2$_1$-type structure. We studied Fe$_2$CoAl Heusler alloy and found that XA ordered structure was much more stable than the regular L2$_1$ structure [26, 27]. The comprehensive theoretical reports are available on L2$_1$ ordered Co$_2$FeAl Heusler alloy [28-34]. However, a proper understanding of the site preferences of 3d atoms in their respective structures (L2$_1$ and XA) is yet to be explored.

The fact is that the accuracy of the density functional calculation is strongly dependent upon the choices of the exchange-correlation function. There are many rundles in Jacob-ladder scheme related to the improvement of correlation to meet an arbitrary level of accuracy [35]. However, a higher number of rundles is always more challenging in computation. Firstly, rundle is known as a local-density approximation (LDA) [36, 37] in which the exchange-correlation (XC) is dependent only on the local density n(**r**). Secondly, the generalized gradient approximation (GGA) [38, 39] depends on the gradient of the density. Generally, GGA is considered more accurate than LDA in the past studies on 3d, 4d and 5d transition metals [40]. Only LDA and GGA for the exchange and correlation potential are not sufficient to describe the complete electronic structure and magnetic behavior of some full Heusler alloys like Co$_2$FeSi [41, 42]. Such strongly correlated systems containing atoms with d and f shells can be treated by adding an on-site Coulomb interaction (U) term by means of GGA+U approach [43, 44]. It is known that the on-site Coulomb interaction (U) may affect the electronic structures and magnetism of the system drastically and exhibit a finite density of states in one spin channel at Fermi energy (E$_F$) [45]. Hence, the effect of electron-electron correlation is expected in Co$_2$FeAl Heusler alloys.

In this paper, the following issues will be addressed:
(1) The literature predicted that Co$_2$FeAl (CFA) Heusler alloy would show the half-metallicity in cubic L2$_1$ (space group (SG): Fm$\bar{3}$m) structure. However, CFA has been experimentally synthesized in cubic A2 and B2 structure. There are no reports on the structural stability of CFA. Hence, we have done a comprehensive study upon the structural stability of CFA in two different cubic symmetries: L2$_1$ under Fm$\bar{3}$m SG (number 225) and inverse XA under F$\bar{4}$3m SG (number 216). Further, we have explored the effects of atomic ordering in their Wyckoff sites: A (0, 0, 0), B (0.25, 0.25), C (0.5, 0.5, 0.5) and D (0.75, 0.75, 0.75) on the ground state properties like half-metallicity, spin polarization and magnetism.
(2) The present literature says that Slater Pauling (SP) rule is a necessary condition along with the exhibiting of an integer magnetic moment to be a perfect half metal. To understand, whether it shows one to one relationship with the electronic structures, we have studied it in details.
(3) GGA is not sufficient to predict the complete physical properties and one has to deal with the exchange and correlation effects present in 3d systems, therefore, we have also focused on U effect upon the ground state properties for all the possible structures.

From the detailed analysis, we found that the ferromagnetic CFA was the most stable in XA-II phase even in comparison with the L2$_1$ phase and became half-metallic within the GGA+U scheme. Hence, the structural dependent studies are crucial and we have addressed in this paper. Our results reveal that CFA is a half-metal not only in L2$_1$ phase but also in the XA phase. This theoretical study will definitely provide guidance for the experimental synthesize of the efficient $Co_2FeAl$ Alloys.

## 2. Computational methods

First-principle calculations have been performed using the Wien2k code [46] which is based on the full-potential linearized augmented plane wave (FPLAPW) method within the density functional theory [47]. To deal with the exchange and correlation among the localized d electrons, generalized gradient approximation (GGA) of Perdew-Burke-Ernzerhof (PBE) was employed in our calculation [48]. The effect of on-site Coulomb interaction (U) in the electronic and magnetic properties was also employed. The effective Coulomb exchange parameter $U_{eff}$ were represented by U-J, where U and J are the Coulomb and exchange parameters, respectively [31]. We kept J value fixed to 0 eV, and U was varied from 0.3 to 1.4 eV; therefore, $U_{eff}$ were equaled to U in our calculation. The maximum value of l ($l_{max}$) for the expansion of wave function in spherical harmonics inside the atomic sphere was restricted to $l_{max}$ = 10. The wave function in the interstitial region was expanded in plane waves with a cutoff of $R_{MT}K_{max}$ = 7, where $R_{MT}$ represents the atomic radii and $K_{max}$ is the largest k vector. The electronic and magnetic properties were studied at their optimized lattice constants ($a_{opt.}$). The charge density was expanded in the interstitial region up to $G_{max}$ = 12 (a.u.$^{-1}$). The crystal structures of $Co_2FeAl$ were generated using the software XCrysDen [49].

## 3. Results and discussions

### 3.1 Structural properties

The full-Heusler ($X_2YZ$ type) alloys are found not only in $L2_1$ or $Cu_2MnAl$ prototype (space group: Fm$\bar{3}$m), but also in XA or $Hg_2CuTi$ prototype (space group: F$\bar{4}$3m) structure. The cubic $L2_1$ structure consists of four interpenetrating fcc sublattices [50]: two of which are equally occupied by X atoms at Wyckoff position 8c (0.25, 0.25, 0.25) and the Y and Z atoms occupy the positions at 4a (0, 0, 0) and 4b (0.5, 0.5, 0.5), respectively (see Fig. 1(a)). Y and Z atoms occupy alternately and hence the centered simple cubic (SC) lattice form a CsCl type superstructure. The 4a and 4b positions have $O_h$ (Octahedral) symmetry in contrary to the 8c position, which has $T_d$ (tetrahedral) symmetry. On the other hand, the XA structure has only $T_d$ symmetry. All four positions: 4a (0, 0, 0), 4b (0.25, 0.25, 0.25), 4c (0.5, 0.5, 0.5) and 4d (0.75, 0.75, 0.75) are adopted with $T_d$ symmetry and $O_h$ symmetry is absent here. The conventional $L2_1$ and inverse XA structures are shown in Fig. 1, where all atoms are occupied in their available Wyckoff sites A (0, 0, 0), B (0.25, 0.25, 0.25), C (0.5, 0.5, 0.5) and D (0.75, 0.75, 0.75) along their body diagonal of the unit cell. In $L2_1$ structure, Co atoms occupy B (0.25, 0.25, 0.25) and D (0.75, 0.75, 0.75) sites, and Fe/Al atoms occupy A (0, 0, 0) and C (0.5, 0.5, 0.5) sites, respectively (see Fig.1 (a)). In XA structure, both inequivalent Co atoms (as labelled Co1/Co2) occupy A (0, 0, 0) and B (0.25, 0.25, 0.25) sites and Fe/Al occupy at C (0.5, 0.5, 0.5) and D (0.75, 0.75, 0.75) sites (see Fig. 1 (b)). Further, we alter the atomic ordering with respect to their Wyckoff sites to make various structures such as L2$_1$-I, L2$_1$-II, XA-I, and XA-II following the scheme as shown in Table 1. Generally, the XA structure is observed in $X_2YZ$ type full Heusler alloys only when the Y atom has a higher atomic number than that of the X atom (from the same period) [51], which is not the case here. Furthermore, it is also known from the site preference rule (SPR) [52, 53] that the element with higher valence (here, cobalt) prefers C (0.5, 0.5, 0.5) site and the element with lower valence (here, Iron) tends to enter at A (0, 0, 0) and B (0.25, 0.25, 0.25) Wyckoff sites. The main group element Al usually prefers D (0.75, 0.75, 0.75) site; in this situation, the alloy tends to form the XA ($Hg_2CuTi$ prototype) structure. On the other hand, when the element with fewer valance electrons tends to occupy the B (0.25, 0.25, 0.25) site, the alloy forms $L2_1$ ($Cu_2MnAl$ prototype) structure. A similar theoretical study on $Fe_2CoGa$ alloy reported that the inverse ($Hg_2CuTi$ prototype) structure is preferable [54]. Heusler alloys (HAs) do not always follow the conventional site preference rule (SPR). To clarify it and to study their ground state properties like equilibrium lattice constants ($a_0$), equilibrium volume ($V_0$), Bulk modulus B, and its pressure derivative B´, we have performed the lattice optimizations for all the structures. The total energy difference ($E_{tot}$ –$E_0$) versus volume curves, which were obtained after fitting with Birch-Murnagham equation [55] have been shown in Fig. 2.

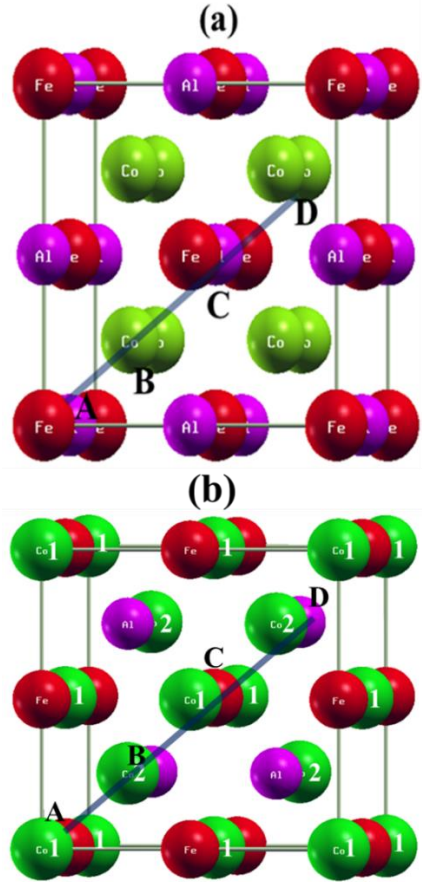

**Fig. 1.** Crystal structures of Co$_2$FeAl (CFA) alloy in (a) conventional L2$_1$ (Cu$_2$MnAl type) and (b) inverse XA (Hg$_2$CuTi type) phase.

**Table 1:** Structural order of regular L2$_1$ (Cu$_2$MnAl) and inverse XA (Hg$_2$CuTi) types CFA, where A, B, C and D represent the available Wyckoff sites: (0, 0, 0), (0.25, 0.25, 0.25), (0.5, 0.5, 0.5) and (0.75, 0.75, 0.75) along the body diagonal, respectively.

| Structure | Wyckoff sites | | | |
|---|---|---|---|---|
| | A | B | C | D |
| L2$_1$-I | Al | Co | Fe | Co |
| L2$_1$-II | Fe | Co | Al | Co |
| XA-I | Co1 | Co2 | Fe | Al |
| XA-II | Co1 | Fe | Co2 | Al |

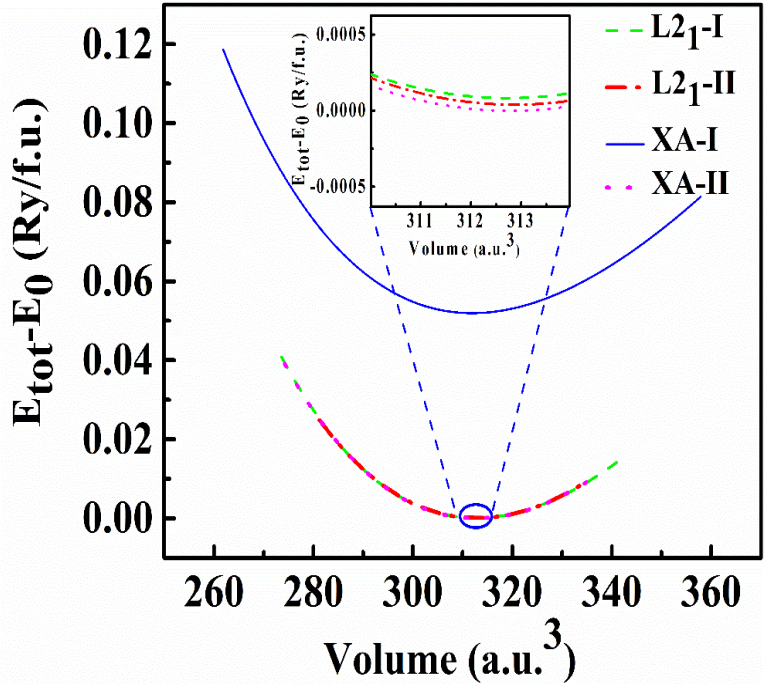

**Fig. 2.** (Color online) Total energy difference ΔE ($E_{tot}$-$E_0$) as a function of the unit cell volume of ferromagnetic $Co_2FeAl$ alloy under $L2_1$-I, $L2_1$-II, XA-I, and XA-II structures. The inset shows a magnified plot of a region near the equilibrium volume $V_0$.

**Table 2:** The calculated optimized lattice parameter $a_0$ (Å), bulk modulus B (GPa) and its pressure derivative B´ of the $Co_2FeAl$ alloy in $L2_1$ ($Cu_2MnAl$) and XA ($Hg_2CuTi$) types under different

atomic arrangements (see Table. 1). The other experimental and theoretical results are shown in brackets for the comparison.

| Parameter | Cu$_2$MnAl (L2$_1$-type) | | Hg$_2$CuTi (XA- type) | |
|---|---|---|---|---|
| | L2$_1$-I | L2$_1$-II | XA-I | XA-II |
| Eq. lattice constants a$_0$ (Å) | 5.70 (5.70)* | 5.70 | 5.69 | 5.70 |
| Bulk modulus B (GPa) | 190.64 (190.19)* | 194.79 | 169.97 | 193.05 |
| Derivative of Bulk modulus (B´) | 4.60 (4.55)* | 3.45 | 5.17 | 4.15 |
| Exp. lattice constants (Å) | 5.72 [20]** (Our experiment), 5.73** [56] (Others) | | | |

* Ref. [57]

**At T=300 K

The fitted parameters: a$_0$, V$_0$, B and B´ are listed in Table. 2. On a comparison of the total energy difference (E$_{tot}$ –E$_0$) versus volume curves, it is clear that the CFA prefers energetically the XA-II ordered structure (but not the conventional L2$_1$ structure) and thus it is more stable than the other structures with an optimized lattice parameter "a$_0$" of 5.70 Å at the equilibrium volume (V$_0$). This optimized lattice parameter (a$_0$) value is comparable with the experimental value shown in Table.2. Moreover, it has shown an opposite tendency to occupy the available Wyckoff sites than those predicted by site preference rule [52, 53]. Hence, our results related to the phase stability of CFA alloy can be taken as a counter-example of the conventional site preference rule (SPR).

### 3.2 Electronic properties within GGA approximation

3.2.1 Density of states

The total and atom specific density of states (DOS) per electron volt (eV) at their respective optimized lattice parameters "a$_0$" are shown in Fig. 3 (a, b) for L2$_1$-I and L2$_1$-II and in Fig. 4 (a, b) for XA-I and XA-II structures. The insets show the magnified plots near the Fermi energy (E$_F$). The comprehensive reports on the conventional L2$_1$ ordered Co$_2$FeAl alloy are available in the literature [28-34, 58, 59], but our results may address the predictability of the device compatibility under conventional L2$_1$ and inverse XA structures and this is important for the device performance. It is clear from Figs. 3 & 4 that the valence band region is split into two parts: lower valance band, which is below -6 eV, and upper valance band from -5.5 eV to the Fermi level (E$_F$). The lower region of the valance band is due to the s-states of Al, which are well separated from the upper valance band states, which consist of p-states of Al and d-states of the two transition metals (Co/Fe). The strong hybridization between 3d-states of Co and Fe atoms has observed in the upper valance band region, which also determines the width and shape of the energy gap in the minority channel. The conduction band above $E_F$ is dominated by d states of Co and Fe atoms. Our result reveals that CFA alloy is a half-metallic ferromagnet under $L2_1$-I structure (see in Fig. 3(a)) and hence forming a net magnetic moment of 5.0 $\mu_B$/f.u per unit cell (see Table. 3).

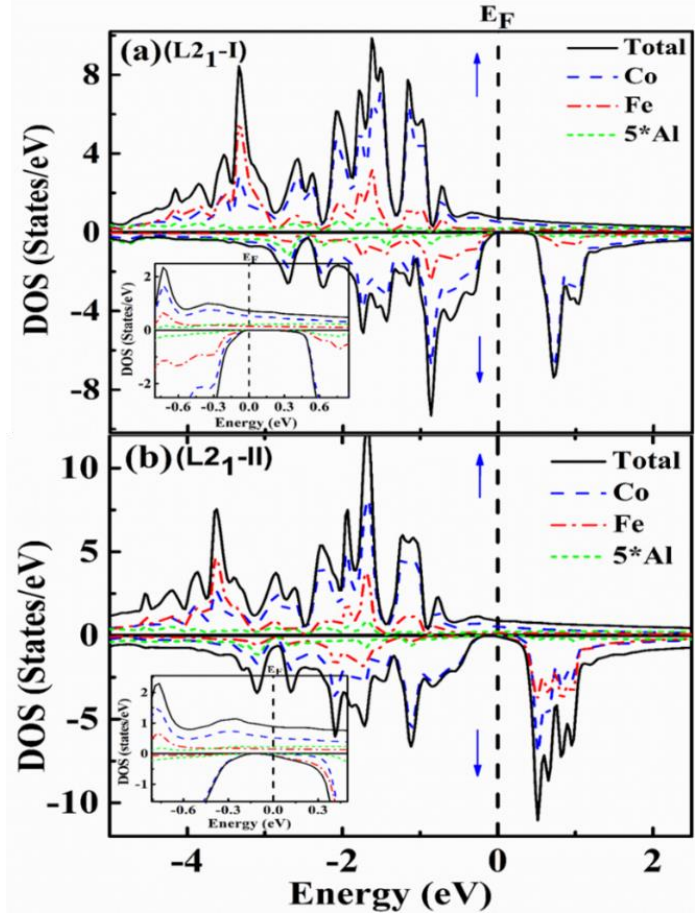

**Fig. 3:** (color online) Plots of total and element-specific density of states (DOS) of $Co_2FeAl$ (CFA) in conventional (a) $L2_1$-I and (b) $L2_1$-II structures. The insets show the magnified plots near Fermi energy ($E_F$). For better visibility, the DOS of the Al atom is enhanced by a factor of 5. The blue ↑ (up arrow) and ↓ (down arrow) represent the up spin channel and down spin channel, respectively.

The calculated total density of states (DOS) exhibit an energy-gap ($E_g\downarrow$) of about 0.2 eV at Fermi energy ($E_F$) for the case of minority spin electrons and conducting due to gapless for the majority electrons; thus 100% spin polarization is observed at Fermi energy ($E_F$). The energy-gap in minority bands is attributed to the Co-Fe interaction, which is the strongest bonding interaction here and hence, bonding and antibonding states are formed due to the covalent hybridization among the transition metals which also determine the position of Fermi energy ($E_F$) [60].

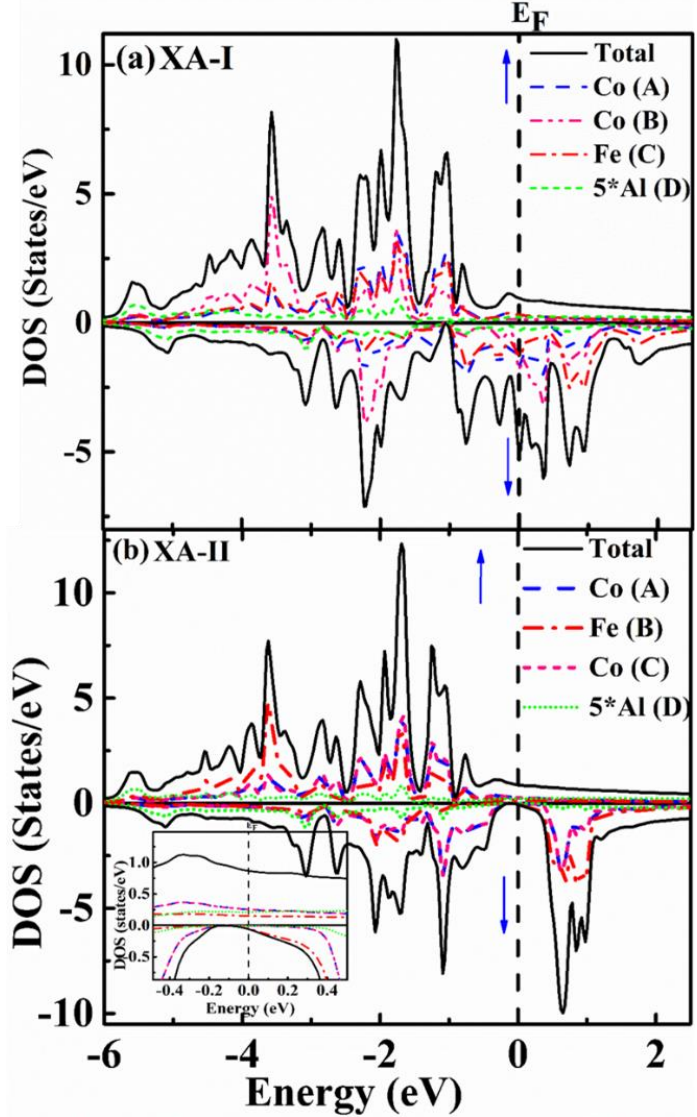

**Fig. 4.** (a, b) Plots of total and element-specific density of states (DOS) of $Co_2FeAl$ (CFA) in XA-I and XA-II structures under inverse XA ($Hg_2CuTi$) prototype. The insets show the magnified plots near the Fermi energy ($E_F$). For better visibility, the DOS of the Al atom is enhanced by a factor of 5. The blue ↑ (up arrow) and ↓ (down arrow) represent the up spin channel and down spin channel, respectively.

The plots of DOS in Fig. 3(b) and Fig. 4(b) also exhibit a gap-like feature (or pseudo-gap) in the spin-down channel but Fermi level ($E_F$) slightly falls away from the gap (see the insets of all figures). Hence, they behave as metals and therefore, the calculated spin polarization (P) has been reduced (see Table. 3). However, the total magnetic moments for these structures are an integer as expected to be 5.0 $\mu_B$/f.u. according to the Slater-Pauling rule [21]. This may be due to that the GGA functional is not able to provide the accurate electronic structure of $Co_2FeAl$. In Fig. 4 (a), the high peak at $E_F$ is mainly due to the cobalt (Co) and Fe $e_g$ states. This high peak is responsible for the instability of inverse XA-I phase (see Fig. 2). As can be seen from the Table. 3, the total magnetic moment is far away from the integer value and hence a very small spin polarization is observed for this structure as compared with the others. Only conventional L2$_1$-I ordered CFA is a half-metal (100% P) at E$_F$, within GGA scheme.

3.2.2 Band Structures

We will discuss here the spin-polarized band structures of Co$_2$FeAl (CFA) alloy in L2$_1$, L2$_1$-II, XA-I, and XA-II phases. In the case of L2$_1$-I type CFA (see Fig. 5(a)); the majority (spin-up) bands are strongly metallic in nature as they are crossing the Fermi level (E$_F$), while minority (spin down) bands show semiconducting nature around the E$_F$.

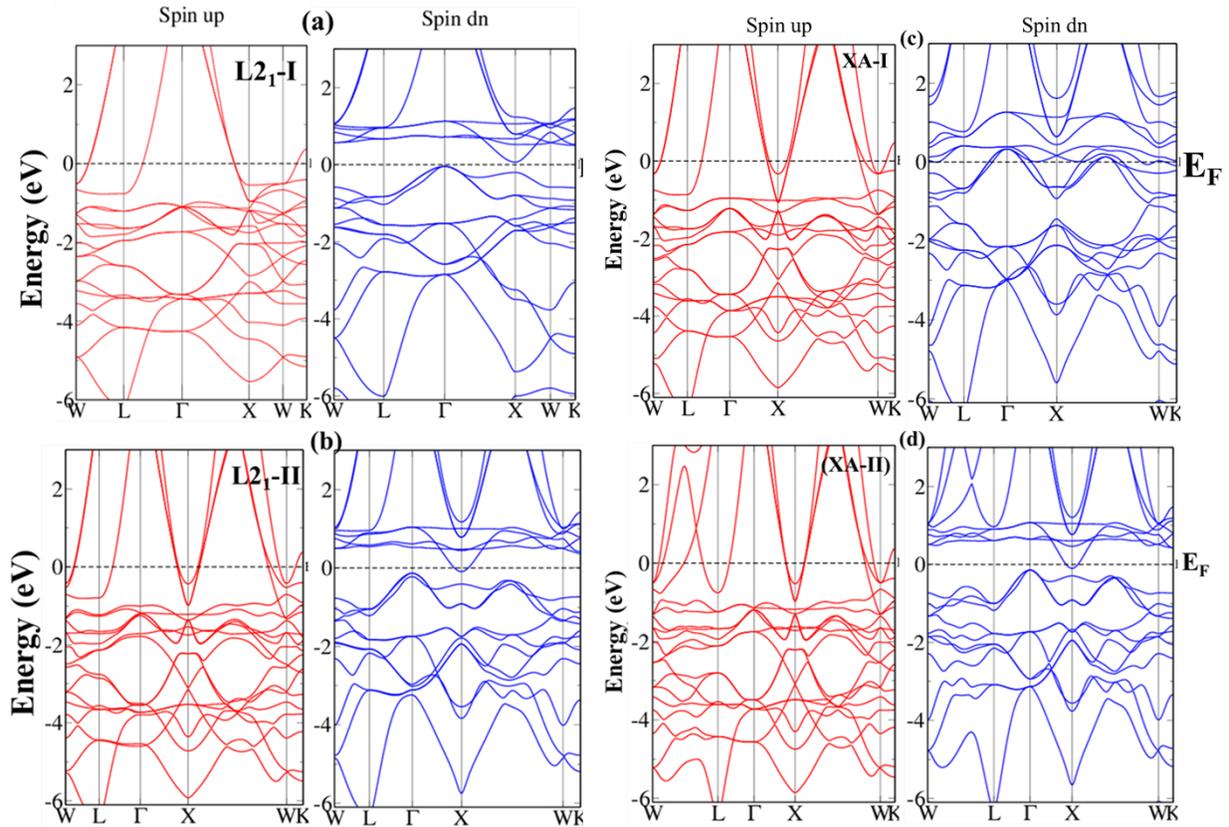

**Fig. 5.** The spin-polarized band structures of Co$_2$FeAl in L2$_1$-I, L2$_1$-II, XA-I, and XA-II phases are plotted in (a, b, c, d), respectively, for the spin-up (red curve) and spin-down (blue curve) electrons within GGA scheme. The dotted line represents the Fermi level (E$_F$).

It can be clearly seen that in the minority spin channel, the valence band maximum (VBM) is located in Γ at -0.046 eV and the conduction band minimum (CBM) is located in X at 0.066 eV and hence, the spin-down indirect (Γ-X) band-gap at around the Fermi level is found to be of about 0.112 eV for Co$_2$FeAl. This band gap is consistent with the previous report by B. Fadila et al. [57]. XA-I type CFA exhibits a metallic nature (see Fig. 5(c)). The band structures of L2$_1$-II and XA-II looks similar and in the spin-down band, an electron-like pocket is observed at X point and also a gap-like feature is observed. Band structure clearly indicates the crossing of a single band in minority spin channel. Therefore, this material seems to be a semimetal rather than a half-metal [61] and results a reduced spin polarization (SP) (see table. 3). Finally, we can conclude that $L2_1$-I type CFA retains its half-metallic nature within GGA scheme, in contrary to the $L2_1$-II, XA-I and XA-II phases of CFA, which can also be seen from DOS plots in Fig. 3&4.

3.2.3 Magnetic properties

It is well known that Co2- based Heusler alloys rigorously follow Slater Pauling (SP) rule [21, 62], which is frequently used to predict their total magnetic moments that linearly scales with the total number of valence electrons. It is also believed that this is a necessary condition to exhibit half-metallicity [57]. The calculated total and atom specific magnetic moments for Co$_2$FeAl alloys under their equilibrium lattice constants ($a_0$) are shown in Table. 3. We have also added the spin-orbit coupling effect in our calculations and the corresponding results are shown in the parenthesis of Table. 3. The results show that the magnetic moment of Co$_2$FeAl is contributed from Co and Fe atoms. Al has a local spin moment of – 0.05 $\mu_B$ and practically no significant contribution in the global moment of CFA.

**Table. 3:** The calculated total magnetic moments ($M_{Tot}^{Cal}$), atom specific magnetic moments ($M_{Co}$, $M_{Fe}$, and $M_{Al}$), reported magnetic moments ($M_{Tot}^{Rep}$), the spin polarization P (%) and Slater-Pauling value ($M_{sp}$) are listed for different structures. Magnetic moments with spin-orbit (SO) coupling are also shown in parenthesis.

| Structure | $M_{Co}$ ($\mu_B$) | $M_{Fe}$ ($\mu_B$) | $M_{Al}$ ($\mu_B$) | $M_{Tot}^{Cal}$ ($\mu_B$) | $M_{Tot}^{Rep}$ ($\mu_B$) | $M_{Sp}$ ($\mu_B$) | P% | Nature |
|---|---|---|---|---|---|---|---|---|
| $L2_1$-I | 1.21 (1.23)* | 2.81 (2.82)* | -0.05 (-0.05)* | 5.00 (5.02)* | 5.08 [28], 4.99[29] | **5.0** | 100 | Half-metal |
| $L2_1$-II | 1.23 (1.21)* | 2.78 (2.80)* | -0.05 (-0.05)* | 4.99 (5.01)* | - | | 85 | Metal |
| XA-I | Co1/Co2 = 1.06(1.06)*/ 1.81(1.82)* | 2.17 (2.19)* | -0.05 (-0.05)* | 4.77 (4.78)* | - | | 68 | Metal |
| XA-II | Co1/Co2 = 1.23(1.22)*/ 1.23(1.22)* | 2.78 (2.74)* | -0.05 (-0.05)* | 4.99 (5.01)* | - | | 87 | Metal |

*Spin-orbit (SO) magnetic moments

The inclusion of Spin-orbit (SO) effect does not change the global moment of CFA significantly in any structures; however, a slight change in partial magnetic moments of the atoms are noticed. Further, it is interesting to note here that the three structures: $L2_1$-I, $L2_1$-II, and XA-II, show the integer magnetic moments (i.e. 5.0 $\mu_B$) approximately, which is consistent with the Slater-Pauling value but only $L2_1$-I structure is a half-metal (100% P) (see Table. 3). Therefore, we employ GGA+U scheme to understand whether the addition of correlation can resolve the inconsistency between the theoretical magnetic moments and electronic structures. Particularly, we will focus on the effects of U in the most stable structure i.e. XA-II phase of the CFA.

**3.3 Effect of on-site Coulomb interaction (U) on electronic properties**

3.3.1 Density of states

From the GGA results of $Co_2FeAl$ Heusler alloy in its $L2_1$-I, $L2_1$-II, XA-I, and XA-II structures, it is seen that only $L2_1$-I structure exhibits a half-metallicity at Fermi energy ($E_F$) and the others are ferromagnetic metals. The existence of finite spin-down states at Fermi energy ($E_F$) may be attributed to the strong exchange-correlation of 3d electrons present in our system. From previous studies on Heusler alloys, it is well known that 3d bands are less dispersive and hence the addition of on-site Coulomb interaction (U) may affect the electronic and magnetic properties of such alloys [63]. Therefore, we further study the effect of U via the GGA+U method, varying the Coulomb potential from 0.34 eV to 1.4 eV for all structures. We have carefully chosen the range of U to study the electronic structures of CFA. We have also performed GGA+U calculations using the previously computed values of 4.22 eV and 4.35 eV at Co and Fe sites [64], respectively, The results are shown in Fig. 7 for comparison. From Fig. 6 (a, e, i), it is clearly visible that the Fermi level ($E_F$) in case of $L2_1$-I structure is exactly in the middle of the spin-down energy gap ($E_g\downarrow$) at U= 0.34 eV, and slightly shifted toward left edge of the gap when U is increased from 0.34 eV to 1.4 eV. The DOS plot at U = 1.4 eV looks similar to the DOS of $L2_1$-I with only GGA scheme (see Fig. 3(a)). Moreover, the total magnetic moment is not affected much by U, but the Fe moment is increased from 2.83 to 2.94 $\mu_B$ (see Table.4). However, Co moment remains the same at U = 0.34 eV to 1.4 eV. Half-metallicity is not affected for any value of U and the respective spin-down energy gap ($E_g\downarrow$) is increased from 0.2 eV to 0.6 eV at U = 0.34 to 1.4 eV indicating the strong covalent hybridization between the 3d atoms.

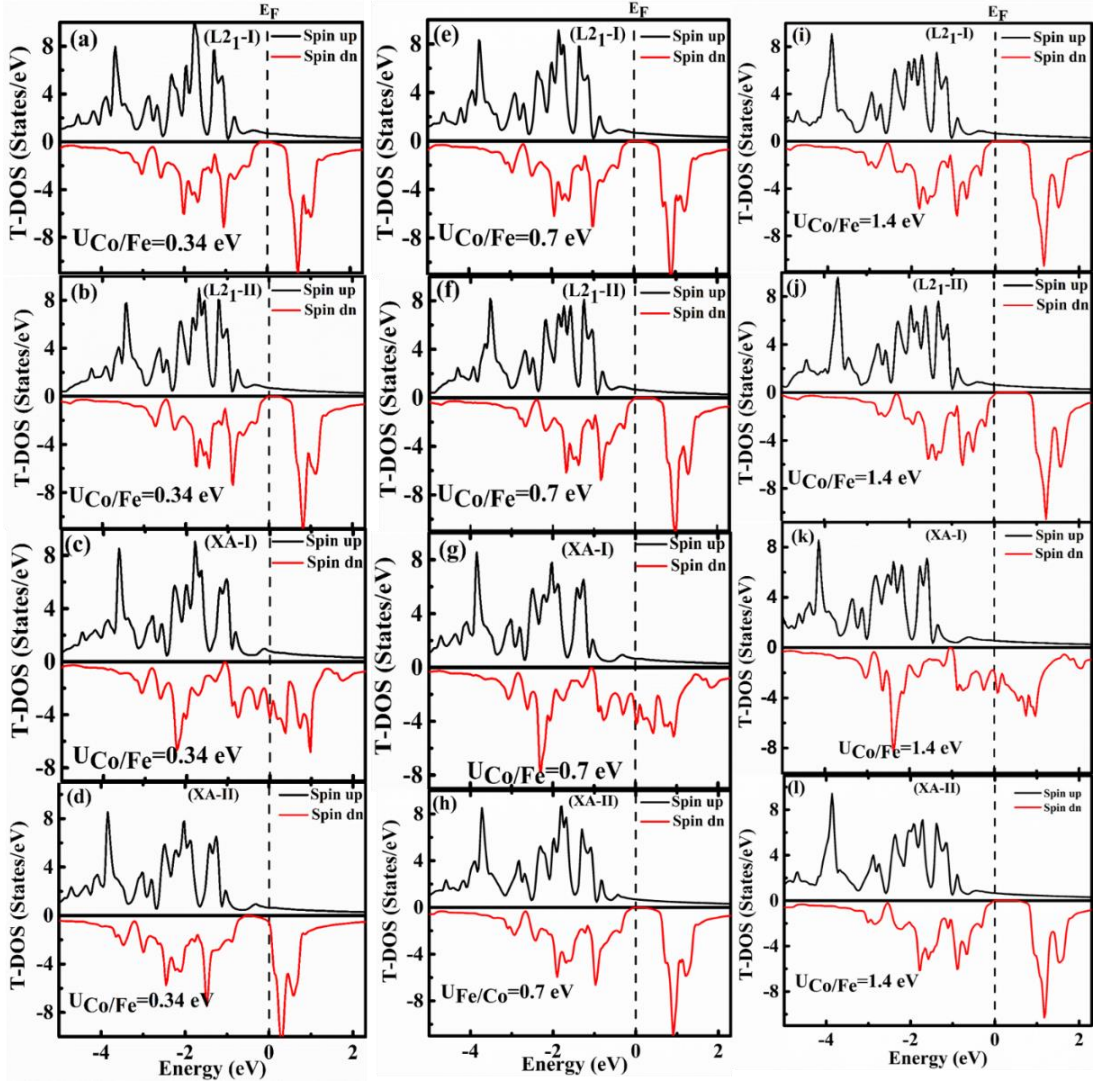

**Fig. 6.** (a-l) (Color online) Plots of the total density of states (T-DOS) in the presence of the on-site Coulomb potential (U). The black and red curves correspond to spin up and spin down DOS, respectively. The U value was varied from 0.34 eV to 1.4 eV and mentioned in all plots; the dotted vertical lines represent the Fermi level ($E_F$).

In $L2_1$-II structure, half-metallicity (100% P) is restored from the metallic nature (for a comparison, see Fig. 3(b)) in the application of U at 0.34 eV and retained up to 0.7 eV. The total magnetic moments of Co and Fe are increased continuously, but the total magnetic moment per cell remains approximately integer (see Table.4). The half-metallicity is destroyed at U=1.4 eV (see Fig. 6(b, f, j)). The gap-width ($E_g\downarrow$) is increased from 0.2 to 0.4 eV at U=0.34 to 0.7 eV.

**Table. 4:** The calculated total ($M_{Tot}^{Cal}$) and elemental magnetic moments ($M_{Co}$, $M_{Fe}$, and $M_{Al}$), the spin polarization P (%) and spin-down band gap $E_g\downarrow$ are listed.

| U (eV) | Structure | $M_{Co}$ ($\mu_B$) | $M_{Fe}$ ($\mu_B$) | $M_{Al}$ ($\mu_B$) | $M_{Tot}^{Cal}$ ($\mu_B$) | P% | $E_g\downarrow$ (eV) |
|---|---|---|---|---|---|---|---|
| 0.34 | L2$_1$-I | 1.22 | 2.83 | -0.06 | 4.99 | 100 | 0.20 |
|  | L2$_1$-II | 1.23 | 2.85 | -0.06 | 4.99 | 100 | 0.20 |
|  | XA-I | Co1/Co2 = 1.15/1.87 | 2.08 | -0.05 | 4.83 | 68 | - |
|  | XA-II | Co1/Co2 = 1.22/1.22 | 2.82 | -0.06 | 4.99 | 8 | - |
| 0.7 | L2$_1$-I | 1.22 | 2.87 | -0.07 | 4.99 | 100 | 0.32 |
|  | L2$_1$-II | 1.24 | 2.90 | -0.07 | 5.00 | 100 | 0.40 |
|  | XA-I | Co1/Co2 = 1.24/1.93 | 2.24 | -0.05 | 5.12 | 69 | - |
|  | XA-II | Co1/Co2 = 1.22/1.22 | 2.87 | -0.07 | 4.99 | 100 | 0.28 |
| 1.4 | L2$_1$-I | 1.22 | 2.94 | -0.09 | 4.99 | 100 | 0.60 |
|  | L2$_1$-II | 1.25 | 2.97 | -0.08 | 5.01 | 51 | - |
|  | XA-I | Co1/Co2 = 1.35/2.01 | 2.43 | -0.06 | 5.47 | 55 | - |
|  | XA-II | Co1/Co2 = 1.22/1.22 | 2.93 | -0.09 | 4.99 | 100 | 0.53 |

Half-metallicity is not observed in XA-I structure (see Fig. 6 (c, g, k)) since the total magnetic moment per cell is very much deviated from the Slater-Pauling rule and hence the spin polarization (P) is also reduced (see Table.4). The general shape of DOS is not affected on the application of U, which is very similar to the DOS obtained from GGA results (for comparison see Fig. 4 (a)). It can be seen from Fig. 6 (d, h, l) that XA-II phase of CFA alloy restores half-metallicity from its metallic nature at U = 0.7 to 1.4 eV, but not at the lower value of U= 0.34 eV. The total magnetic moment of Co1 and Co2 remains the same, but the magnetic moment of Fe is constantly increased from 2.82 to 2.93 $\mu_B$ (see Table.4) at U = 0.34 to 1.4 eV. Moreover, the total magnetic moments per cell are not affected by any value of U and consistent with the Slater-Pauling rule.

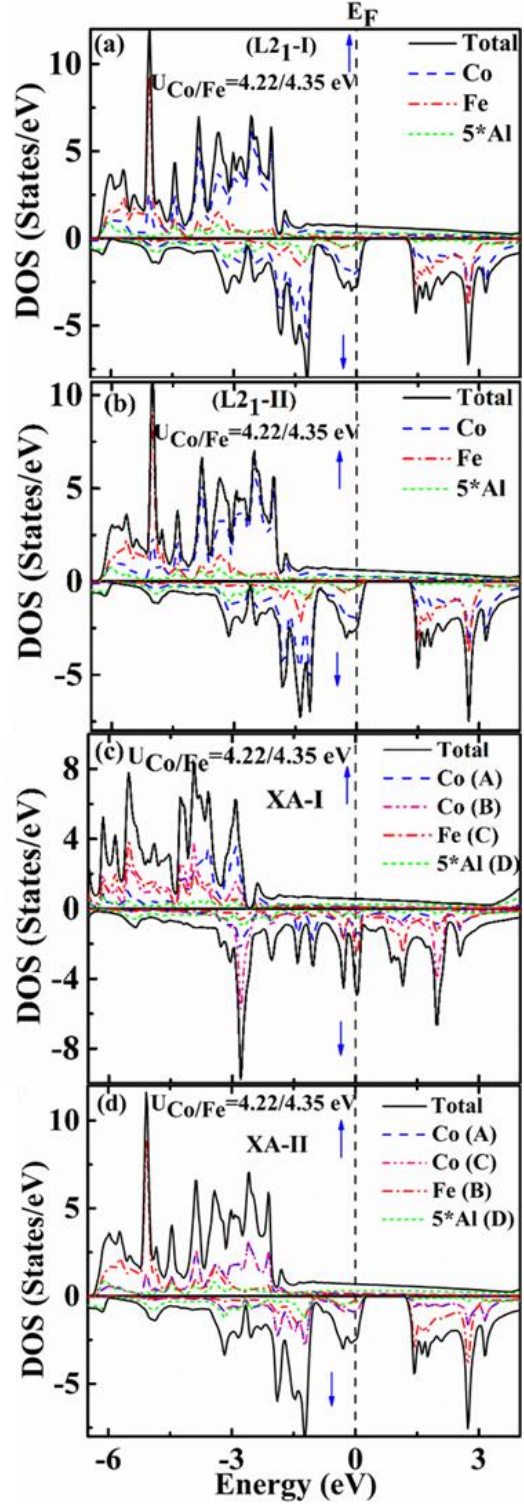

**Fig. 7.** (a, b, c, d) Plots of the total and element-specific density of states (DOS) of $Co_2FeAl$ (CFA) in $L2_1$-I, $L2_1$-II, XA-I, and XA-II structures in the presence of U = 4.22 and 4.35 eV for Co and Fe atoms, respectively. For better visibility, the DOS of the Al atom is enhanced by a factor of 5. The blue ↑ (up) and ↓ (down) arrows represent the up spin channel and down the spin channel, respectively.

The band gap (E$_g\downarrow$) is increased from 0.28 to 0.53 at U=0.7 to 1.4 eV. We have also plotted the total and atom specific density of states calculating at U=4.22 and 4.35 eV [64] for Co$_2$FeAl alloy in Fig. 7. The Co$_2$FeAl alloy exhibits the metallic nature for all structures at those values of U. Furthermore, it can be seen from the Table. 5 that Co and Fe moments are constantly varied for all structures and very much deviated from the Slater-Pauling value of 5.0 µ$_B$/f.u.; hence the spin polarization (P) is also decreased. From the above discussion, it is clear that the electronic and magnetic properties are not only dependent on L2$_1$ and XA ordering of the atoms in their respective structures, but also on the choices of U values. GGA results show the underestimation of the energy gaps (E$_g\downarrow$) and half-metallicity compared to those of GGA+U results. GGA+U results clearly show that the metallic ground state of L2$_1$-II and XA-II phases is changed to half-metallic ground state and the spin-down band gaps are improved. The electrical conductivity and neutron diffraction experiments may confirm the predictions.

**Table. 5:** The calculated total ($M_{Tot}^{Cal}$) and element-specific magnetic moments (M$_{Co}$, M$_{Fe}$, and M$_{Al}$) and the calculated spin polarization P (%) are listed for different structures.

| U(Co/Fe) (eV) | Structure | M$_{Co}$ (µ$_B$) | M$_{Fe}$ (µ$_B$) | M$_{Al}$ (µ$_B$) | $M_{Tot}^{Cal}$ (µ$_B$) | P% | Nature |
|---|---|---|---|---|---|---|---|
| 4.22/4.35 | L2$_1$-I | 1.50 | 3.18 | -0.14 | 5.66 | 59 | Metal |
| | L2$_1$-II | 1.53 | 4.00 | -0.12 | 5.67 | 61 | Metal |
| | XA-I | Co1/Co2 = 1.68, 2.09 | 3.05 | -0.10 | 6.40 | 79 | Metal |
| | XA-II | Co1/Co2 = 1.52/ 1.52 | 3.19 | -0.14 | 5.71 | 57 | Metal |

These experiments could directly probe the nature of the magnetic coupling between Co and Fe moments and electronic transport behavior. In literature, no such data for L2$_1$ and XA phases of Co$_2$FeAl alloy are available. However, our results will help to understand the role of on-site Coulomb potential (U) between the 3d electrons of Co and Fe atoms in Co$_2$FeAl alloy.

3.3.2 Band structures

We present the results of band structure calculations on the most stable phase XA-II of CFA to explore the insight of the electronic structures. The spin-polarized band structures of Co$_2$FeAl (in XA-II phase) have been calculated at different U values from 0.34 eV to 1.4 eV and the results are shown in Fig. 8 (a, b, c). From Fig. 8(a), Spin up bands exhibit metallic nature due to the crossing of conduction bands at Fermi level (E$_F$). On the other hand, in spin-down bands, one can clearly see the Fermi level of a single band crossing at X point and hence XA-II phase is found to be semimetal. When U is increased to 0.7 and 1.4 eV, XA-II phase of CFA restores its half-metallicity from its metallic nature (see Fig. 8 (b, c)) with a spin down indirect ($\Gamma$-X) band gap of about 0.101 eV and 0.120 eV at the U values of 0.7 and 1.4 eV, respectively. The results of bands structures are in a close agreement with the results of DOS plots as discussed above.

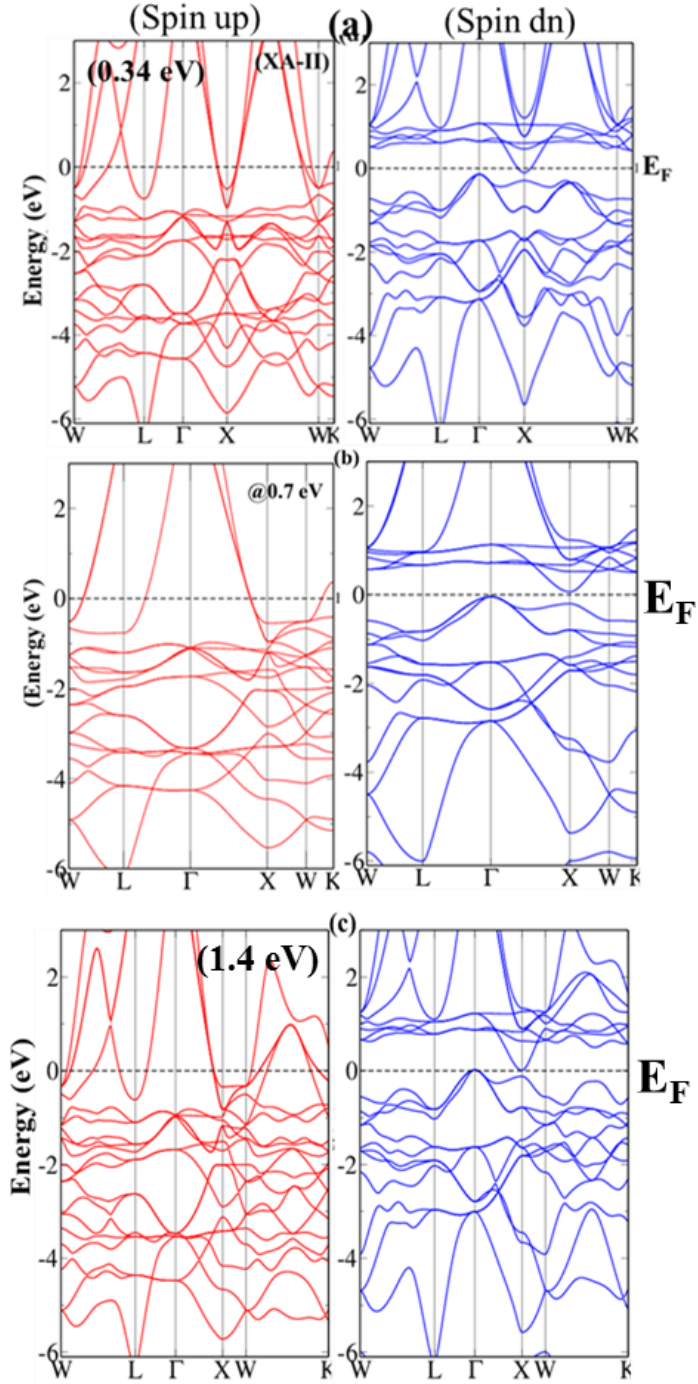

**Fig. 8:** The band structures of $Co_2FeAl$ in XA-II phase in the presence of the on-site Coulomb potential (U) are shown at (a) 0.34 eV, (b) 0.7eV and (c) 1.4 eV. The red and blue curves correspond to the spin-up and the spin-down bands, respectively. The dotted horizontal line represents the Fermi level $E_F$.

## 4. Conclusion

We have pursued an in-depth study of the effects of atomic ordering in their respective Wyckoff sites on phase stability, half-metallicity and magnetic properties under conventional L2$_1$ and inverse XA structures. The GGA results revealed that Co$_2$FeAl alloy became half-metallic-ferromagnet only in L2$_1$-I type structure with a 100% spin polarization at Fermi energy ($E_F$). However, the other structures showed the metallic nature, but their spin polarization values were significantly high at Fermi energy ($E_F$). The XA-II structure was found to be energetically most preferable compared to the others with a spin polarization of around 84% at $E_F$. GGA+U calculations presented the more accurate results than the GGA considering a careful choice of U. Some structures restored their half-metallicity in the presence of the on-site Coulomb interaction (U). Further, we have proved that the atomic occupation at their Wyckoff sites in CFA was decisive in determining their electronic and magnetic properties. The results revealed that the Slater-Pauling rule, which was commonly used to predict the new half-metallic Heusler alloys, would still be the necessary condition. Hence, our work is highly instructive for the design of efficient CFA alloys for the application in spintronics.

## 5. Acknowledgments

Aquil Ahmad sincerely acknowledges the University Grant Commission (UGC) Delhi, MHRD Delhi, India for providing fellowship for Ph.D. work. A. K. Das acknowledges the financial support of DST, India (project no. EMR/2014/001026). Our fruitful discussion with Dr. Monodeep Chakraborty is highly acknowledged. We also acknowledge our departmental computational facility, IIT Kharagpur, India